\documentclass{PoS}

\usepackage{mathrsfs}
\usepackage{amssymb}
\usepackage{amsmath}
\usepackage{graphicx}
\usepackage{amsfonts,amsbsy}
\usepackage{color}
\usepackage{nicefrac}
\usepackage{xcolor}
\usepackage{comment}

\newcommand{\mbf}{\mathbf}

\newcommand{\mrm}{\mathrm}

\newcommand{\Tr}{\mrm{Tr}}
\newcommand{\HTL}{\mrm{HTL}}

\newcommand{\Q}{Q}

\newcommand{\fig}{Fig.~}

\newcommand{\se}{Sec.~}

\newcommand{\re}{Ref.~}
\newcommand{\res}{Refs.~}

\newcommand{\ud}{\mathrm{d}}

\title{Understanding the dynamics of field theories far from equilibrium}

\ShortTitle{Understanding the dynamics of field theories far from equilibrium}

\author{\speaker{Kirill Boguslavski}\thanks{The author is grateful to J.\ Berges, A.\ Chatrchyan, J.\ Jaeckel, A.\ Kurkela, T.\ Lappi, J.\ Peuron, A.\ Pi\~{n}eiro Orioli, S.\ Schlichting, R.\ Venugopalan and R.\ Walz for collaboration on different projects presented here, and P.\ Arnold, M.~A.\ Escobedo, T.\ Gasenzer, M.\ Laine, A.\ Mazeliauskas, J.\ Pawlowski, A.\ Rebhan, P.\ Romatschke, K.\ Rummukainen, C.~A.\ Salgado and M.\ Strickland for valuable discussions. 
He would also like to thank the organizers for the interesting conference and for the invitation to give this talk.
This work is supported by the European Research Council under grant no.~ERC-2015-CoG-681707.}\\
        Department of Physics, P.O.~Box 35, 40014 University of Jyv\"{a}skyl\"{a}, Finland\\
        E-mail: \email{kirill.boguslavski@jyu.fi}}

\abstract{In recent years, there have been important advances in understanding the far-from-equilibrium dynamics in different physical systems. In ultra-relativistic heavy-ion collisions, the combination of different methods led to the development of a weak-coupling description of the early-time dynamics. The numerical observation of a classical universal attractor played a crucial role for this. Such attractors, also known as non-thermal fixed points (NTFPs), have been now predicted for different scalar and gauge theories. An important universal NTFP emerges in scalar theories modeling ultra-cold atoms, inflation or dark matter, and its scaling properties have been recently observed in an ultra-cold atom experiment. In this proceeding, recent progress in selected topics of the far-from-equilibrium evolution in these systems will be discussed. A new method to extract the spectral function numerically is a particularly promising tool to better understand their microscopic properties.}

\FullConference{XIII Quark Confinement and the Hadron Spectrum - Confinement2018\\
		31 July - 6 August 2018\\
		Maynooth University, Ireland}

\begin{document}

\section{Introduction}

In recent years there have been important advances in understanding isolated quantum systems in extreme conditions far from equilibrium. Prominent examples include the reheating process in the Early Universe after inflation, the initial stages in ultra-relativistic heavy-ion collisions (URHICs) at large laboratory facilities like the Large Hadron Collider (LHC) at CERN and the Relativistic Heavy Ion Collider (RHIC) at BNL, as well as table-top experiments with ultra-cold quantum gases. Even though the typical energy scales of these systems vastly differ, they can show very similar dynamical phenomena at weak self-couplings $g^2 \ll 1$.

\begin{figure}[t]
	\centering
	\includegraphics[scale=0.55]{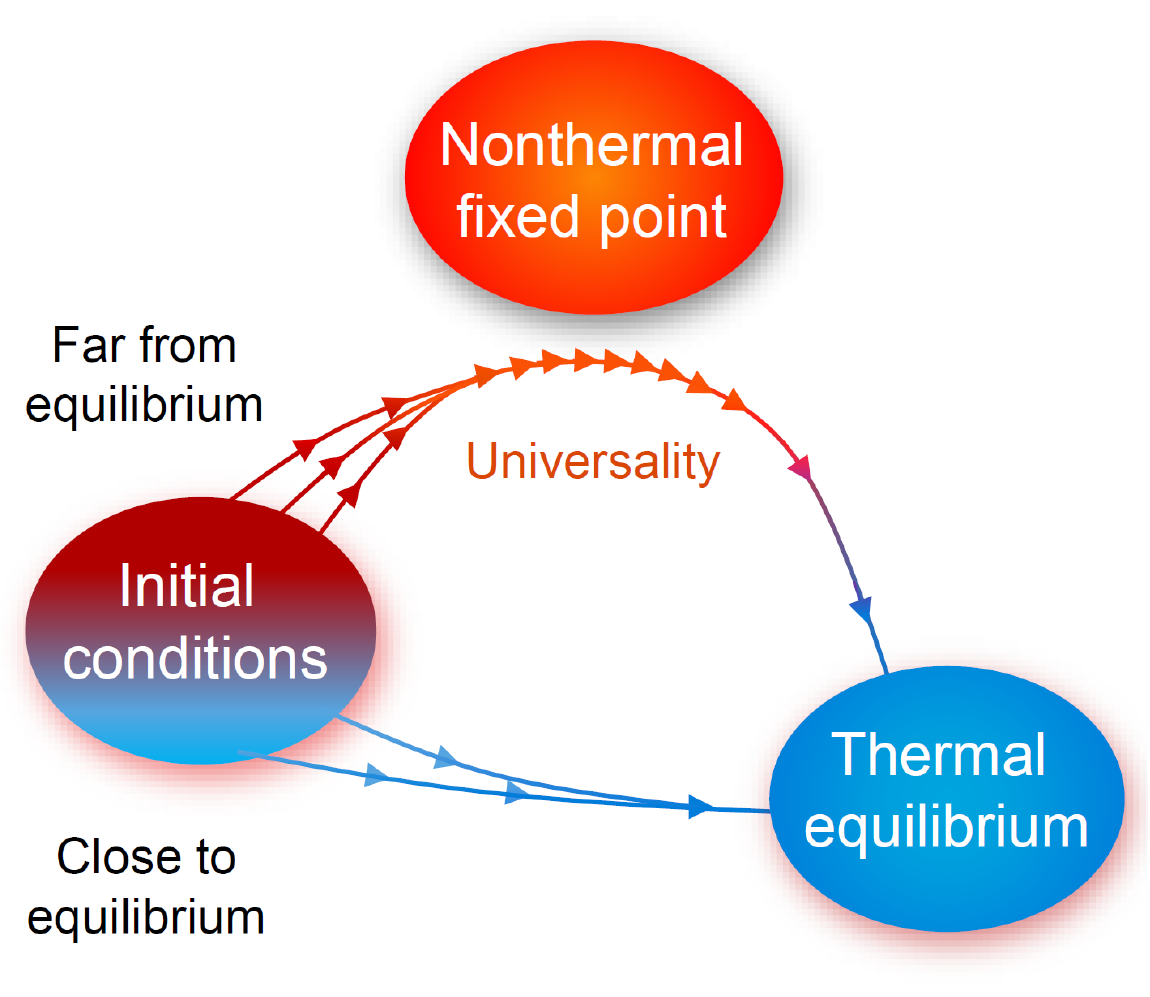}
	\caption{Visualization of different ways to approach thermal equilibrium. Far-from-equilibrium initial conditions like high initial occupation numbers $f \sim 1/g^2$ may lead to an NTFP where the dynamics slows down considerably and the memory of details of the initial state gets lost. Eventually, the system thermalizes, leaving the attractor. (figure developed as an online illustration of the highlighted article \re\cite{Berges:2014bba})}
	\label{fig_NTFP}
\end{figure}

\subsection*{Non-thermal fixed points (NTFPs) and universality classes far from equilibrium}

An important phenomenon emerging far from equilibrium is the concept of a nonthermal fixed point (NTFP) \cite{Berges:2008wm}, which is illustrated in \fig\ref{fig_NTFP}. When the initial conditions are far from thermal equilibrium, as, for instance, with large typical occupation numbers $f \sim 1/g^2$, the system can approach an NTFP during its space-time evolution. 
However, when occupation numbers become $f \sim 1$, quantum effects draw it away and towards thermal equilibrium. NTFPs are far-from-equilibrium attractor states and can be characterized by a self-similar evolution of the single-particle distribution function
\begin{align}
 f(t,p) = (\Q t)^\alpha f_s((\Q t)^\beta p),
 \label{eq_selfsim_iso}
\end{align}
where $t$ is time, $p$ is momentum and $\Q$ is a constant momentum scale. The scaling exponents $\alpha$, $\beta$ and the scaling function $f_s(p)$ are universal, i.e., not sensitive to the details of the original initial conditions \cite{Micha:2004bv,Berges:2013lsa,Kurkela:2012hp}. Therefore, when approaching an NTFP, the system loses its memory of the initial state. An NTFP is often connected to a conservation law 
that relates the scaling exponents. In some systems, like scalar field theories, more than one scaling region can be found. Usually two regions appear, one at low (IR) and another one at high momenta (UV), and correspond to particle number and energy conservation, respectively. The IR region will be discussed in \se\ref{sec_NTFP_atoms_cosmo}.

Such attractor states have been found in different quantum field theories describing physical systems. Important examples involve URHICs, where a discovered NTFP \cite{Berges:2013eia,Berges:2013fga} helped to establish our current understanding of the dynamics at early times in heavy-ion collisions at weak couplings, models of the inflaton during cosmological reheating in the Early Universe \cite{Kofman:1994rk}, where an NTFP in the UV led to the ``Turbulent thermalization'' scenario 
\cite{Micha:2004bv}, and ultra-cold atomic gases \cite{Nowak:2010tm}
where the scaling properties of a universal NTFP in the IR have been predicted in \re\cite{Orioli:2015dxa} and where this and other scaling behavior have been recently observed experimentally in \res\cite{Prufer:2018hto,Erne:2018gmz}.

Universality classes far from equilibrium are an intriguing new concept that connects NTFPs from different theories \cite{Berges:2014bba,Orioli:2015dxa}. Here one uses the set of exponents and scaling function $\alpha$, $\beta$ and $f_s(p)$ of an NTFP to define a universality class. In general, an NTFP depends on the underlying microscopic field theory, its symmetries, the momentum region of interest, conservation laws and the metric. If, however, the scaling properties are the same for different microscopic theories, then the theories are said to lie in the same universality class. Such behavior has been found 
in \cite{Berges:2014bba} between scalar and gauge theories with a longitudinally expanding metric that reflects the geometry in URHICs, and in \cite{Orioli:2015dxa} for different scalar theories that underlie descriptions of ultra-cold atoms and inflation models in cosmology. In \cite{Berges:2015ixa} it was even found that NTFPs of both universality classes emerge in different momentum regions of a scalar system. This provides an intriguing link connecting URHICs, ultra-cold atoms and the Early Universe in a single system.

\subsection*{Methods to study far-from-equilibrium dynamics}

In these systems, the large occupation numbers $f \gg 1$ at relevant modes enable an approximative description of the full quantum field theory in terms of classical field dynamics while preserving the full quantum initial state \cite{Aarts:2001yn}, which is referred to as the classical-statistical approximation (CSA). The CSA is a commonly used method to study far-from-equilibrium phenomena on a real-time lattice and have been used for the systems discussed above. Kinetic theory is another important tool, where one studies the underlying microscopic dynamics in terms of scattering processes of quasiparticles if occupation numbers are not too large $f \lesssim 1/g^2$ and a quasiparticle assumption is valid \cite{Jeon:2004dh}.
When a scale separation between soft and hard typical momentum modes exists, hard thermal loop (HTL) perturbation theory is also applicable \cite{Braaten:1989mz}. Recently, a new method has been developed in \re\cite{Boguslavski:2018beu} that, extending CSA simulations, enables to obtain the spectral function in highly-occupied out-of-equilibrium systems directly in real time, which will be discussed in \se\ref{sec_spec_fct}.

\subsection*{Content and outline}

In \se\ref{sec_phys_sys}, we focus on the emergence of NTFPs and their importance for the respective physical systems. An emphasis will be given to recent progress in the extraction and description of their properties and how they link disparate physical systems. A new numerical method to obtain spectral functions for dynamical systems will be presented in \se\ref{sec_spec_fct}. This introduces a novel tool to study far-from-equilibrium dynamics. In \se\ref{sec_conclusion}, we summarize and discuss some open questions in the considered physical systems and how they can be addressed with the new method.

\section{Non-thermal fixed points in different physical systems}
\label{sec_phys_sys}

We will now discuss important applications of far-from-equilibrium dynamics in different physical systems with large occupation numbers. With Quantum Chromodynamics (QCD), relativistic scalar field theories and non-relativistic scalar field theories, respectively, their underlying quantum field theories are vastly different. Nevertheless, the emergence of NTFPs and universality poses strong links between them.

\subsection{NTFP at initial stages in ultra-relativistic heavy-ion collisions}
\label{sec_IS_URHICs}

In ultra-relativistic heavy-ion collisions (URHICs) as performed at LHC and RHIC a locally equilibrated quark-gluon plasma (QGP) is expected to be formed shortly after the collision. The quarks and gluons forming this state interact via the strong force QCD. Having an ab-initio understanding of the initial stages during which the plasma reaches local equilibrium (or when a description in terms of hydrodynamics becomes valid) is crucial for the phenomenological analysis of heavy-ion collisions. 

The evolution of the plasma to local equilibrium is a multi-stage process. At initial stages, the plasma is far from thermal equilibrium. Using weak-coupling techniques, it is expected to be in the approximately boost-invariant Glasma state right after the collision \cite{Krasnitz:1998ns,Krasnitz:2001qu,Lappi:2003bi}.
This state is, however, unstable because quantum vacuum fluctuations break the approximate boost invariance explicitly. This results in exponential growth of quantum fluctuations for a range of momenta, which are called instabilities \cite{Mrowczynski:1993qm,Rebhan:2004ur,Romatschke:2006nk}.
The growth stops when non-linear interactions of these unstable modes become important \cite{Berges:2012iw,Berges:2012cj}. This leads to a highly occupied mainly gluonic plasma with occupation numbers being $f(\Q) \sim 1/g^2$ for typical momenta $\Q$, where $g^2$ is the gauge self-coupling. 

\begin{figure}[t]
	\centering
	\includegraphics[scale=0.235]{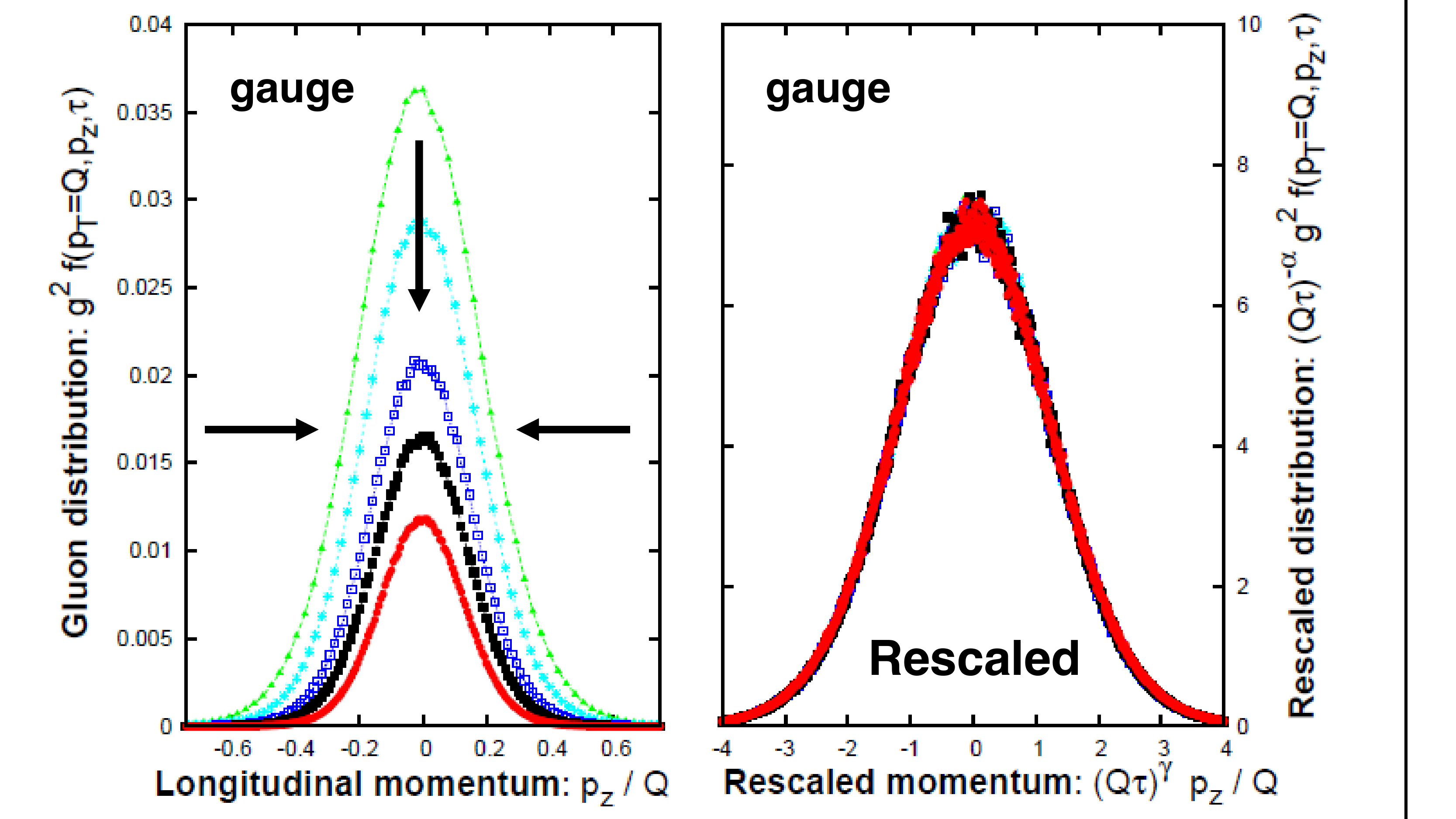}
	\includegraphics[scale=0.47]{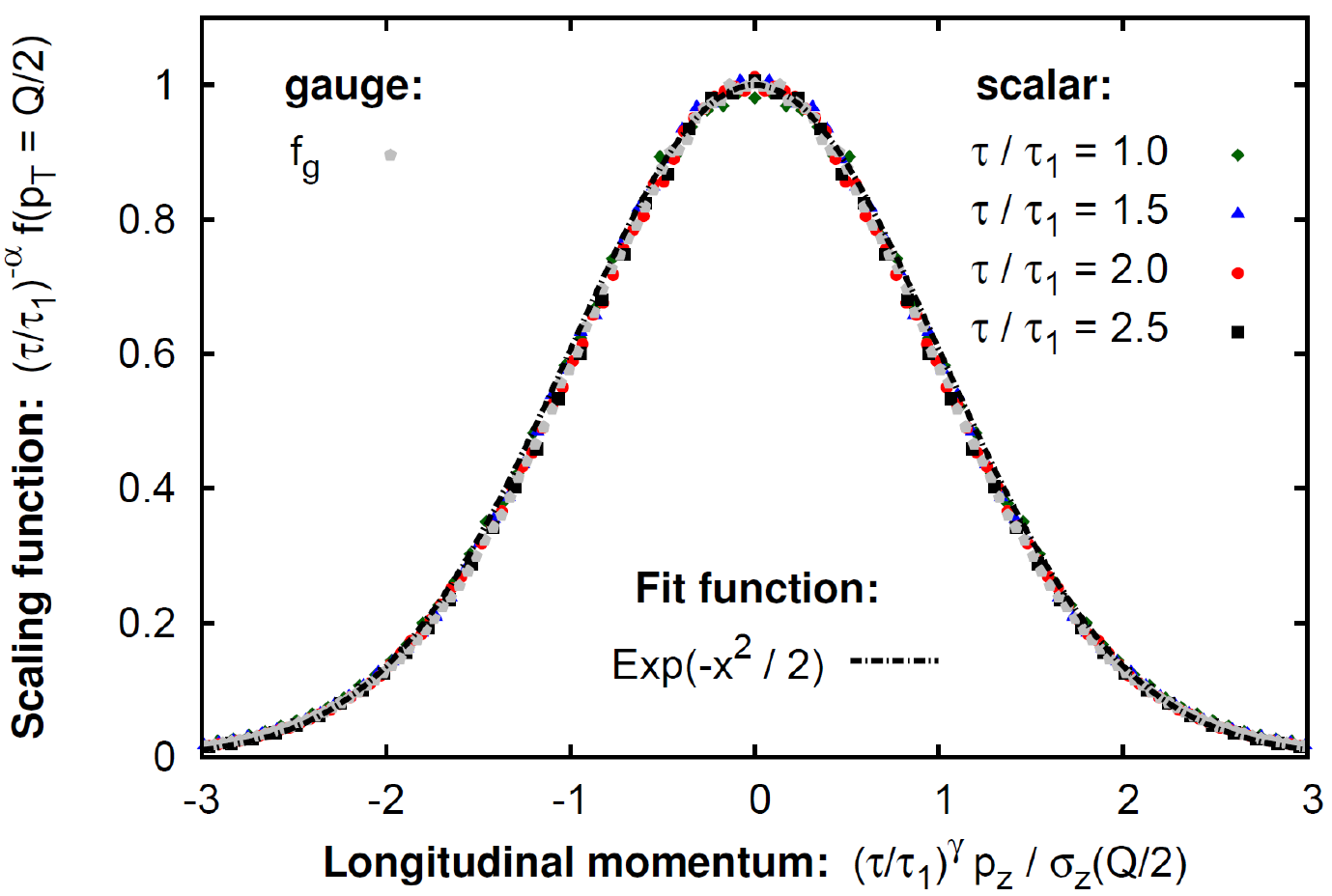}
	\caption{Self-similar behavior of expanding gauge and scalar systems at fixed transverse momentum $p_T$ shown as a function of longitudinal momentum $p_z$ for different times. {\bf Left:} The distribution is shown at different times for a non-Abelian gauge plasma. The arrows indicate that occupation numbers and typical longitudinal momenta decrease. Next to it, the same distributions are shown with rescaled occupation numbers and momenta. (figure taken from \cite{Berges:2013eia}) {\bf Right:} The longitudinal distribution of a scalar system is shown for different times, the longitudinal gauge distribution is also included. Their momenta and occupation numbers have been normalised at the reference time $\tau_1$ to become comparable. Curves at later times $\tau > \tau_1$ have been obtained by rescaled occupation numbers and momenta with power laws in time with the universal exponents \eqref{eq_scalexp_gauge_aniso}. (figure published in \cite{Berges:2014bba} with few changes)}
	\label{fig_gauge_scalar_selfsim}
\end{figure}

With CSA lattice simulations, the subsequent evolution of highly occupied non-Abelian plasmas was studied in \res\cite{Berges:2013eia,Berges:2013fga}. It was shown that for a wide range of initial conditions, the plasma approaches an anisotropic NTFP with\footnote{To account for the high velocities of the colliding nuclei, it is convenient to change time $t$ and spatial coordinate $z$ to eigen-time $\tau$ and rapidity $\eta$. This introduces a longitudinally expanding (Bjorken) metric. As a consequence, the system becomes anisotropic between transverse and longitudinal momenta $p_T$ and $p_z$, respectively.}
\begin{align}
 f(\tau,p_T,p_z) = (\Q \tau)^\alpha f_s((\Q \tau)^\beta p_T, (\Q \tau)^\gamma p_z)
 \label{eq_selfsim_aniso}
\end{align}
and universal scaling exponents
\begin{align}
 \label{eq_scalexp_gauge_aniso}
 \alpha = -\frac{2}{3}\;, \quad \beta = 0\;, \quad \gamma = \frac{1}{3}\;.
\end{align}
The self-similar evolution is demonstrated in the left panel of \fig\ref{fig_gauge_scalar_selfsim} for an $\mrm{SU}(2)$ non-Abelian gauge theory. Because of $\beta = 0$, we only show here that the distribution at fixed $p_T$ is self-similar as a function of $p_z$. While the occupation numbers and typical longitudinal momenta shrink with time as illustrated by the arrows, rescaling them with suitable power laws in time results in a single curve as demonstrated in the adjoining plot marked as ``Rescaled''. This curve corresponds to the universal scaling function $f_s$ that, together with the scaling exponents, characterizes the NTFP. Its functional form turns out to be of Gaussian shape. 

In \re\cite{Berges:2014bba}, the same scaling properties have been found in CSA lattice simulations for $O(N)$ symmetric relativistic scalar theories with quartic interactions and the same longitudinally expanding metric. Despite being entirely different theories, relativistic scalars exhibit a scaling momentum region with the same scaling exponents \eqref{eq_scalexp_gauge_aniso} and the same scaling form $f_s$. This is demonstrated in the right panel of \fig\ref{fig_gauge_scalar_selfsim}. Here the longitudinal scalar distribution at fixed $p_T$ is rescaled for different times using the universal scaling exponents, with all distributions falling onto the same curve. For comparison, a gauge theory curve is also included. This shows that this scaling behavior in longitudinally expanding scalar and gauge theories forms a universality class far from equilibrium. 

The finding of this NTFP in non-Abelian gauge plasmas was a crucial step in the formulation of a consistent picture of the initial stages in URHICs. While different kinetic scenarios had been suggested for the non-thermal plasma, the extracted scaling exponents \eqref{eq_scalexp_gauge_aniso} favored the `bottom-up' scenario \cite{Baier:2000sb} that only requires elastic $2 \leftrightarrow 2$ and inelastic effectively $2 \leftrightarrow 1$ processes and can be formulated within an effective kinetic theory \cite{Arnold:2002zm}. While the bottom-up scenario consists of three stages, the NTFP corresponds to its first stage. Numerical simulations of this kinetic theory confirmed the scaling properties of the NTFP 
and are incorporated in current descriptions of the pre-equilibrium (and pre-hydrodynamic) evolution in URHICs \cite{Kurkela:2015qoa,Kurkela:2018wud}.

\begin{figure}[t]
	\centering
	\includegraphics[scale=0.57]{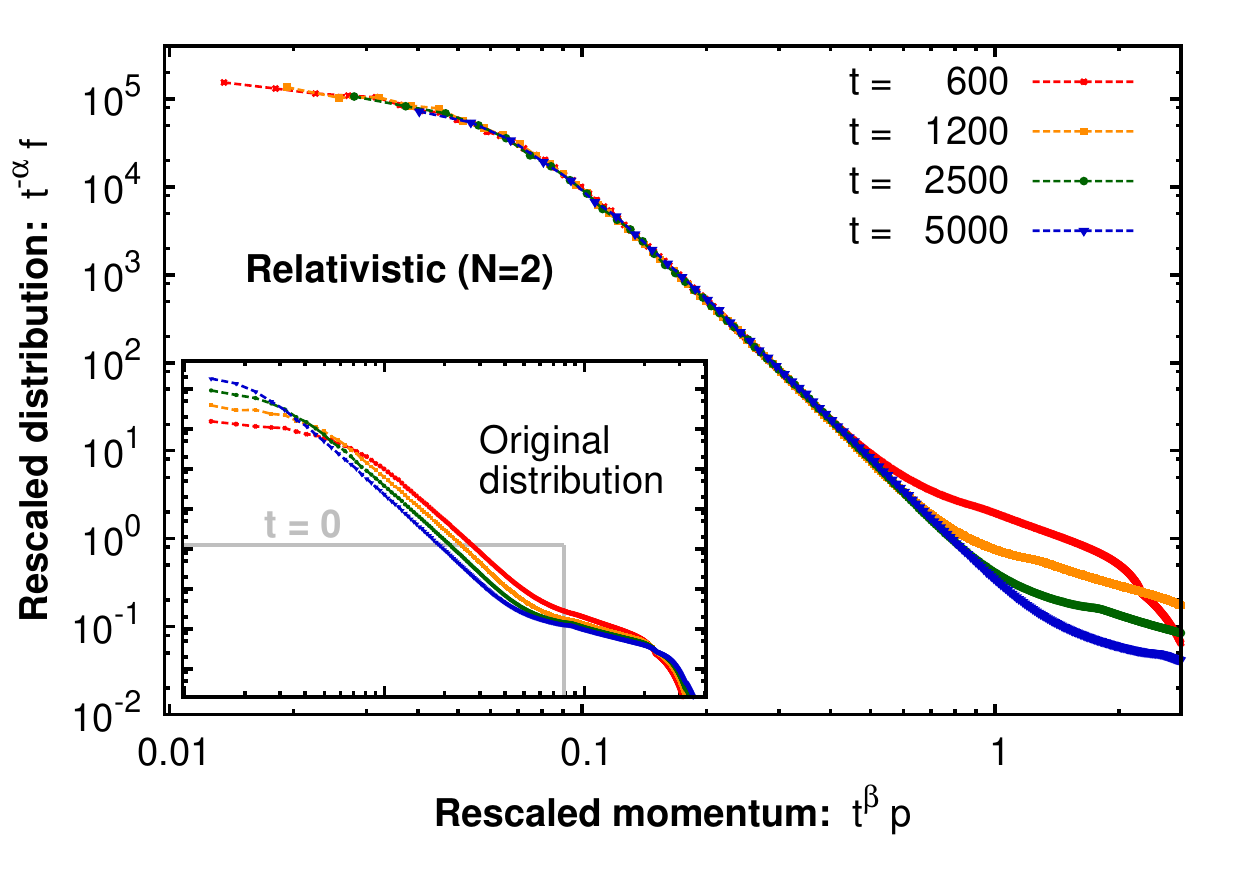}
	\includegraphics[scale=0.21]{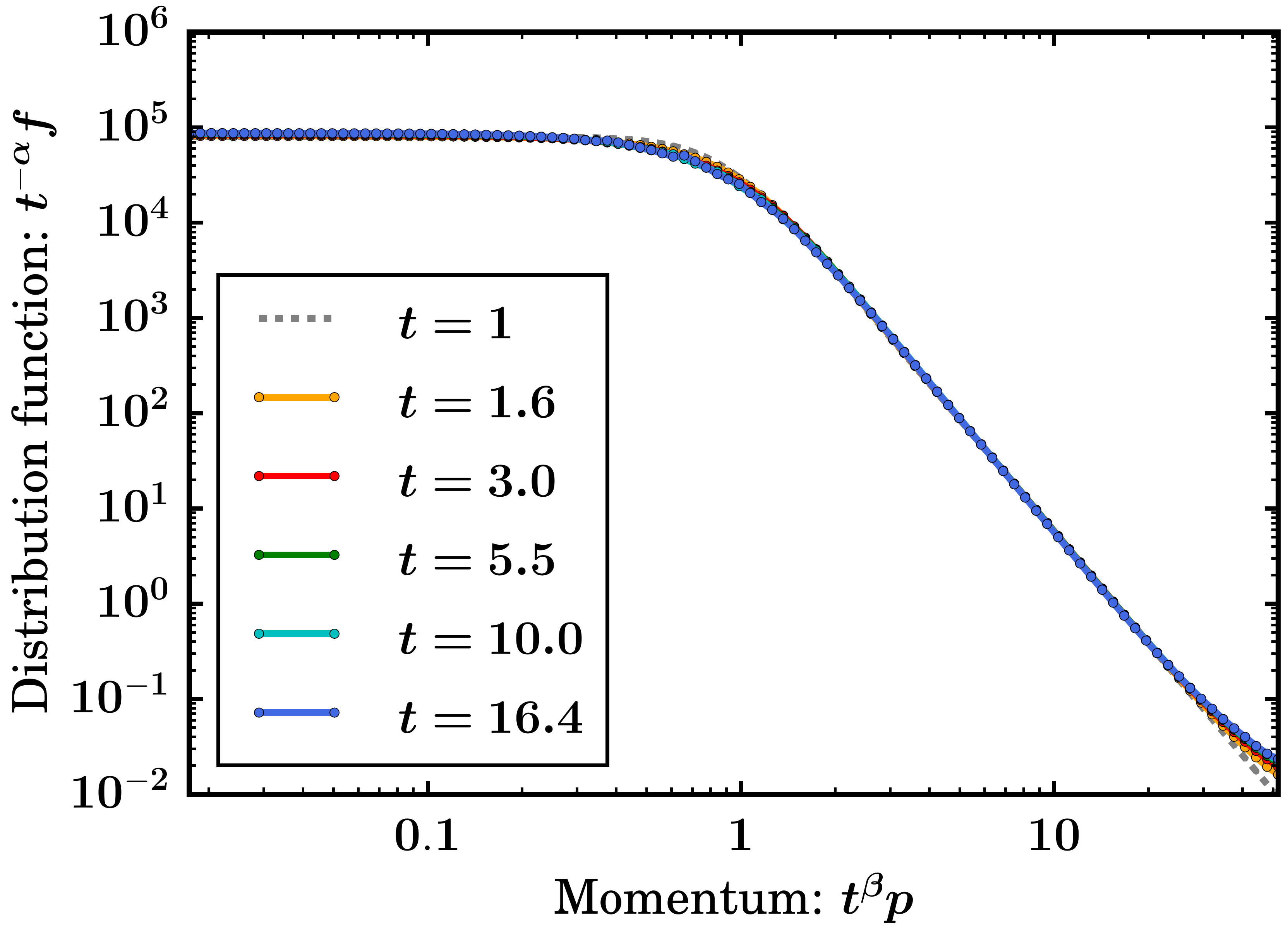}
	\caption{Self-similar evolution of the IR region in scalar systems. Times and momenta are normalized differently in the panels, while the same exponents \eqref{eq_scalexp_IR_scalars} for $d=3$ are used. 
	{\bf Left:} The rescaled distribution function is shown at different times as function of rescaled momentum for a relativistic scalar theory with $N=2$ fields. The inset illustrates the original distribution before rescaling. These data have been computed with CSA numerical simulations. (plot taken from \cite{Orioli:2015dxa}) 
	{\bf Right:} The rescaled distribution is shown at different times for a non-relativistic scalar theory. These curves have been computed with the Large-$N$ kinetic theory. (figure taken from \cite{Walz:2017ffj})}
	\label{fig_scalars_IR_selfsim}
\end{figure}

\subsection{NTFPs with ultra-cold atoms and cosmological models}
\label{sec_NTFP_atoms_cosmo}

Another interesting class of NTFPs emerges in physical systems that can be described by scalar field theories. Prominent examples involve ultra-cold atoms and cosmological models of inflation or of dark matter. There it is possible to construct initial conditions where typical occupation numbers of momentum modes below a characteristic scale $p \lesssim \Q$ are large $f \sim 1/\lambda$ while $\lambda \ll 1$ is weak. In the context of ultra-cold atoms, $\lambda$ corresponds to the diluteness parameter, which is proportional to the quartic self-coupling\footnote{Usually, its symbol is $g$. We changed it to $g^2$ here to be consistent with the common notation for gauge theories.} $g^2$ of a non-relativistic scalar field theory \cite{Nowak:2010tm,Orioli:2015dxa}. For the given examples in cosmology, $\lambda$ is the quartic self-coupling of $O(N)$ symmetric relativistic scalar field theories with a $\lambda (\varphi_a \varphi_a)^2/(24N)$ interaction term in the potential. A situation of large occupation numbers can arise, for instance, after an instability phase during preheating \cite{Micha:2004bv,Kofman:1994rk}.

In the subsequent evolution, the highly occupied systems approach attractors that involve two scaling regions, one at low momenta in the infrared (IR) and one at high momenta for ultraviolet (UV) modes. Each region has a separate set of scaling exponents and functions and evolves self-similarly according to \eqref{eq_selfsim_iso}. While at high momenta, energy density is conserved and transported to the UV \cite{Micha:2004bv}, in the low-momentum region particle density is conserved and transported to lower momenta, adding to the zero-momentum mode \cite{Berges:2008wm,Nowak:2010tm,Orioli:2015dxa}. The scalar attractor hence forms a dual cascade.

In general, every scalar theory can have a separate scaling region. For instance, non-relativistic and relativistic (massless) scalar theories have different scaling properties in the UV \cite{Micha:2004bv}. However, the IR region turns out to be the same for different $N$ and for both relativistic and non-relativistic scalar theories, which has been found in \re\cite{Orioli:2015dxa}. The scaling behavior computed with CSA simulations in that work is shown in the left panel of \fig\ref{fig_scalars_IR_selfsim} for a relativistic scalar theory with $N=2$ fields. While the inset shows the original distribution as a function of momentum at different times, the curves lie on top of each other after rescaling in the main plot. The exponents are to a good approximation given by
\begin{align}
 \alpha = \frac{d}{2}\;, \quad \beta = \frac{1}{2}\,,
 \label{eq_scalexp_IR_scalars}
\end{align}
for $d=3$ spatial dimensions. The scaling function $f_s$ is the same for different $N$ as well as for non-relativistic scalar theory, forming a universality class of scalar theories in the IR.

The properties of this region can be understood within a Large-$N$ kinetic theory 
\begin{equation}
 \frac{\partial f(t,p)}{\partial t} = C[f](t,p)\,,
 \label{eq_Boltzmann_equation}
\end{equation}
with a collision integral $C$ that results from a systematic expansion in $1/N$ to next-to-leading order (NLO) \cite{Berges:2008wm,Orioli:2015dxa,Berges:2001fi,Scheppach:2009wu}. Plugging the scaling ansatz \eqref{eq_selfsim_iso} into the kinetic equation already leads to the scaling exponents \eqref{eq_scalexp_IR_scalars}, as demonstrated in \re\cite{Orioli:2015dxa}. The scaling function $f_s$ has been determined in \re\cite{Walz:2017ffj} by solving the Large-$N$ kinetic theory numerically, which is shown in the right panel of \fig\ref{fig_scalars_IR_selfsim}. The rescaled distribution is shown at different times as a function of rescaled momentum. With time, all curves fall on top of each other, forming the scaling function $f_s$. As for the lattice results shown in the left panel, the lowest momenta describe a rather flat distribution, while higher momenta follow a $p^{-\kappa}$ power law with $\kappa = 4$ for the kinetic theory in $d=3$ dimensions (see also \cite{Chantesana:2018qsb}) and $\kappa \simeq 4 - 4.5$ for the lattice results. We note that the observed scaling can also be described with a recently developed effective low-energy theory \cite{Mikheev:2018adp}. 

Such a scaling region emerges in a variety of different scalar systems. Such include theories with $\phi^6$ self-interaction at sufficiently large occupancies even when the quartic coupling is negative \cite{Berges:2017ldx}, theories in different spatial dimensions
\cite{Gasenzer:2011by,Karl:2016wko,Boguslavski:2018xxx} 
or with a longitudinally expanding metric \cite{Berges:2015ixa}. In the latter two cases, the scaling function or the scaling exponents may receive corrections but can still be well described by the Large-$N$ kinetic theory. Remarkably, in a recent experiment with a spinor Bose gas \cite{Prufer:2018hto}, an NTFP was found with the scaling behavior \eqref{eq_scalexp_IR_scalars} predicted in \cite{Orioli:2015dxa}.


\begin{figure}[t]
	\centering
	\includegraphics[scale=0.34]{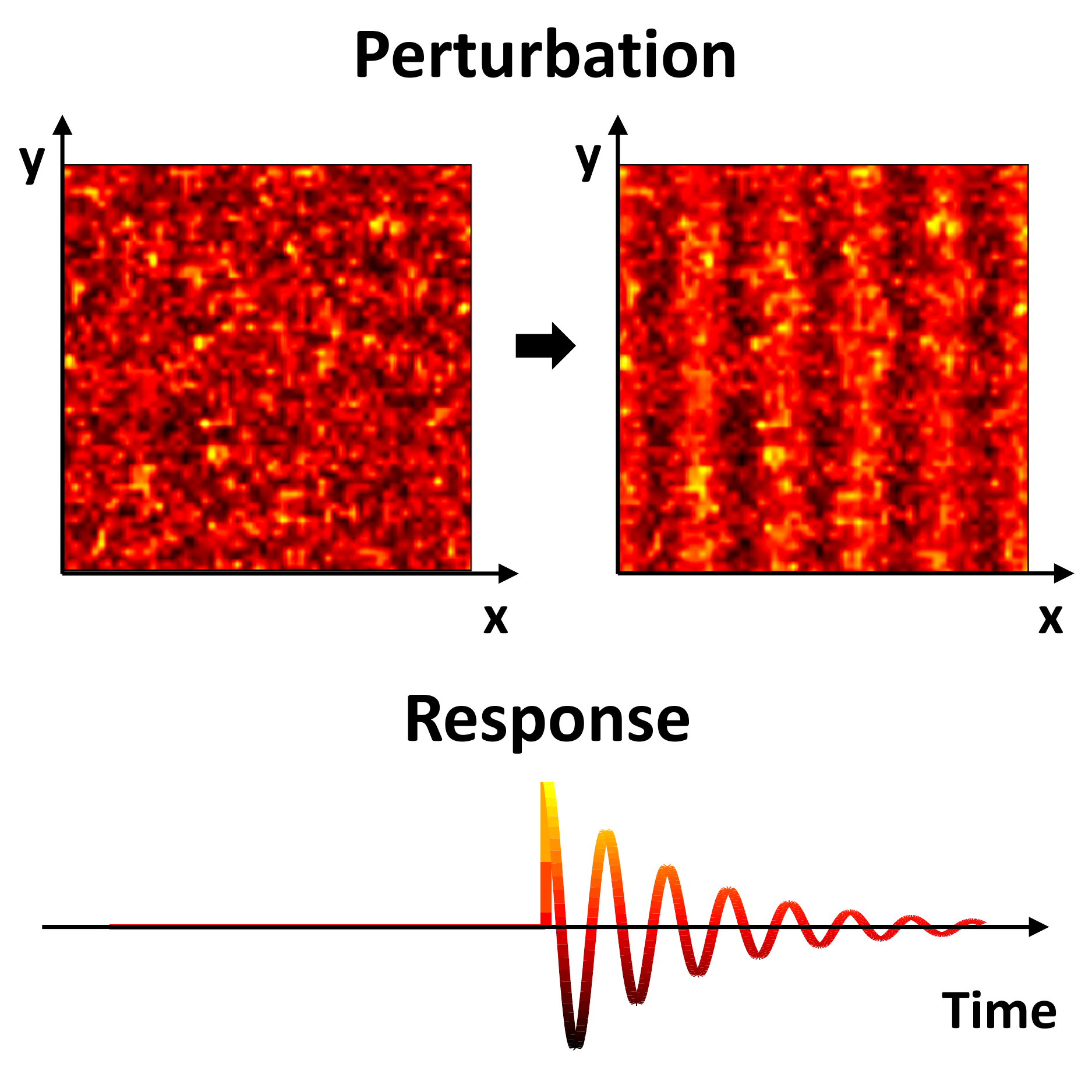}
	\caption{Illustration of a new method to compute spectral functions on a classical lattice, see main text for a description. (figure developed as an online illustration of the highlighted article \re\cite{Boguslavski:2018beu})}
	\label{fig_spectral_principle}
\end{figure}

\section{Next step: Spectral functions}
\label{sec_spec_fct}

While we have mainly discussed the evolution of the distribution function so far, the next step is to study the underlying excitation spectrum that is encoded in the spectral function $\rho(t,\omega,p)$. This time-dependent quantity shows the excitation spectrum in frequency domain for each momentum. Here quasiparticles emerge as Lorentz peaks 
\begin{align}
 \label{eq_lorentz_peak}
 \propto \frac{1}{(\omega - \omega(t,p))^2 + \gamma^2(t,p)}
\end{align}
with the quasiparticles' energy (dispersion relation) $\omega(t,p)$ and inverse life time (damping rate) $\gamma(t,p)$. Additionally, more complicated structures may emerge like Landau damping or extra poles. With spectral functions, non-equilibrium dynamics can be better understood and quasiparticle assumptions underlying kinetic theories can be tested. 

A numerical method has been recently developed in \cite{Boguslavski:2018beu} to compute the spectral function for non-Abelian gauge theories, that uses linearized equations of motion that had been derived before in \cite{Kurkela:2016mhu}. The new method combines linear response theory with CSA lattice simulations. While we will not discuss details of the method here, its working principle is illustrated in \fig\ref{fig_spectral_principle}. In CSA simulations, gauge $A_j(t,\mbf x)$ and chromo-electric fields $E^j(t,\mbf x)$, with $j = 1,2,3$, are discretized on a cubic lattice of size $N_s^3$ with lattice spacing $a_s$ and are evolved by solving classical equations of motion (EOM) of $\mrm{SU}(N_c)$ gauge theory in temporal gauge.\footnote{To be more precise, the gauge covariant link fields $U_j(t,\mbf x) \approx \exp(i g a_s A_j(t,\mbf x))$ are used instead of the gauge fields in the evolution equations. See, e.g., \res\cite{Berges:2013fga,Boguslavski:2018beu} for details.} In the upper left part of the figure, a snapshot of a 2-dimensional slice of the lattice is shown for $\sum_j \Tr(E^j E^j)$, where the trace goes over gauge degrees of freedom. This system is perturbed by an instant source $j$ at time $t'$ with momentum $\mbf p$ that corresponds to a plane wave in $x$ direction, resulting in the snapshot shown in the upper right panel right after the perturbation.\footnote{The remaining gauge freedom is fixed by Coulomb gauge $\left.\partial_j A_j\right|_{t=t'}=0$ right before the perturbation to provide the interpretation of the source as a perturbation at momentum $\mbf p$.} 
Numerically, the response fields caused by the perturbation are separated from the gauge fields $A_j(t,\mbf x) \mapsto A_j(t,\mbf x) + a_j(t,\mbf x)$ and analogously for $E^j$. While the remaining background fields follow the unperturbed classical EOM as before, the response fields $a_j$ and $e^j$ are evolved to later times $t > t'$ by solving linearised equations of motion developed in \cite{Kurkela:2016mhu}. The retarded propagator $G_R$ is extracted from a (classical) average of the response field by use of linear response theory
\begin{align}
 \langle a_j(t,\mbf p) \rangle = G_{R,jk}(t,t',\mbf p)\,j^k(t',\mbf p),
 \label{eq_linresp_GR}
\end{align}
where $a_j(t,\mbf p)$ is the Fourier transformed response field. To reduce computational costs, the source perturbs the system at several momentum modes simultaneously.\footnote{This trick has been developed for non-relativistic scalar theory in \re\cite{PineiroOrioli:2018hst} and adapted to non-Abelian gauge theory in \re\cite{Boguslavski:2018beu}.}
Finally, the spectral function $\rho$ results from
\begin{align}
 G_{R,jk}(t,t',\mbf p) = \theta(t-t')\,\rho_{jk}(t,t',\mbf p)\,.
\end{align}

This technique has been applied to an isotropic non-Abelian gauge theory as a first application. Its NTFP reveals a self-similar evolution \eqref{eq_selfsim_iso} that transports energy density to harder momenta with scaling exponents $\beta = -1/7$ and $\alpha = -4/7$ \cite{Kurkela:2012hp}. 
In this system, a hard scale $\Lambda(t)$ can be defined as the momentum where the energy density $\sim \int \ud^3p\,p\,g^2f(t,p)$ receives its dominating contributions while a mass $m$ can be defined using the hard thermal loop (HTL) formalism \cite{Braaten:1989mz}. The self-similar evolution leads then to $\Lambda(t)/\Q \sim (\Q t)^{1/7}$ and $m(t)/\Q \sim (\Q t)^{-1/7}$, where $\Q$ is a constant momentum scale, such that the scale separation between soft and hard modes increases with time as
\begin{align}
 \Lambda(t) / m(t) \sim (\Q t)^{2/7} \gg 1\,.
\end{align}
Because of this scale separation, the HTL formalism is applicable to this system and the spectral function determined with the new numerical method can be compared to known results from HTL.

\begin{figure}[t]
	\centering
	\includegraphics[scale=0.55]{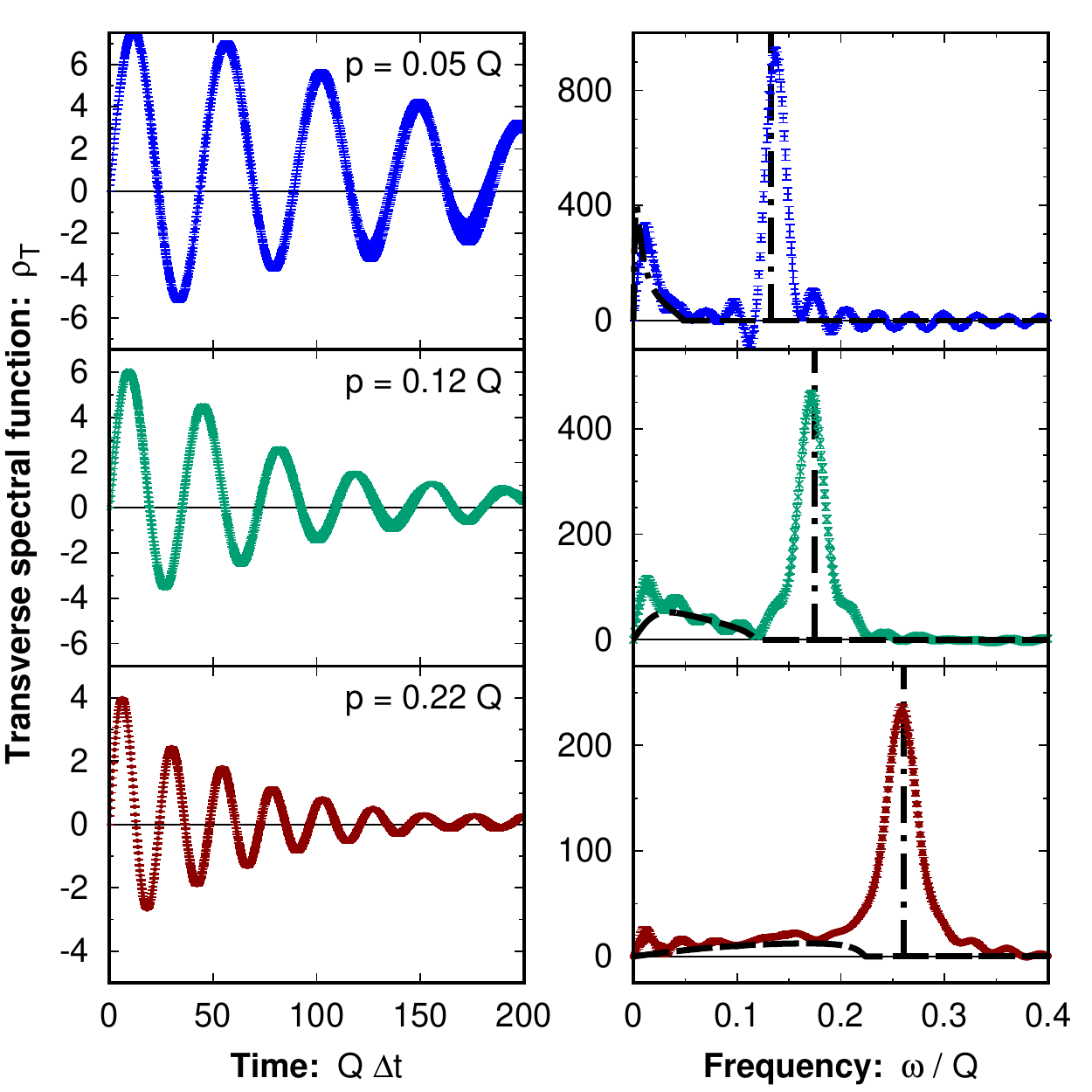}
	\includegraphics[scale=0.63]{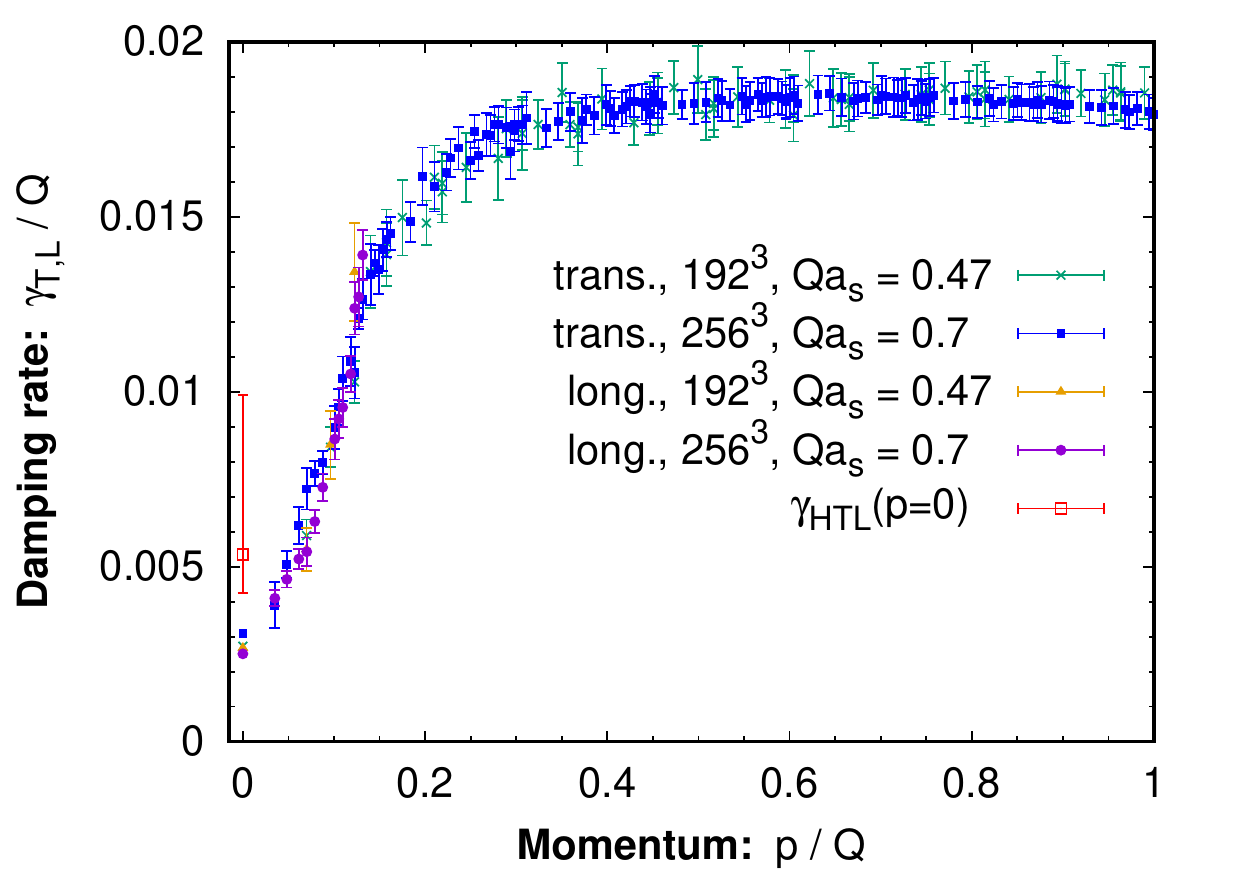}
	\caption{{\bf Left:} Spectral function of transversely polarized gauge bosons at different momenta as a function of time difference $\Delta t$ and frequency $\omega$. Dashed black lines are computed consistently in the HTL formalism at LO. {\bf Right:} The corresponding damping rate (inverse life time) of quasiparticles for both transverse and longitudinally polarized bosons for different lattice discretizations as functions of momentum. For comparison, the corresponding asymptotic mass is $m \approx 0.15\,\Q$. The value $\gamma_{\HTL}(p=0)$ as calculated in the HTL formalism \cite{Braaten:1990it} is also included. (both figures are taken from \cite{Boguslavski:2018beu})}
	\label{fig_spectral}
\end{figure}

The transversely polarized spectral function,\footnote{One can distinguish longitudinal and transverse polarizations $\rho_L$ and $\rho_T$ by projecting $\rho_{jk}$ on the sub-space spanned by $P^L_{jk} = p_j\, p_k^*/p^2$ and on $\delta_{jk} - P^L_{jk}$, respectively.} 
computed as described above, is shown in the left panel of \fig\ref{fig_spectral} for different momentum modes at a fixed time $t$ when the system is close to its NTFP and, thus, evolves self-similarly. In the left part of that panel, $\rho_T$ is plotted as a function of the time difference $\Delta t = t-t'$ with $\Delta t \ll t$ while its Fourier transform with respect to $\Delta t$ is plotted in the right part of the left panel. Additionally, black dashed curves correspond to the spectral function computed in the HTL formalism at LO. 
For each momentum mode, one observes two structures in the frequency domain: a quasiparticle excitation emerging as a Lorentzian peak \eqref{eq_lorentz_peak} and a continuum at low frequencies $|\omega| \leq p$ that can be identified as Landau damping. 
Remarkably, the HTL curves at LO agree well with the Landau cut region and with the position of the peak, i.e., with the extracted dispersion relation. 

The LO HTL curves mainly differ from the data by neglecting the finite width of the peak, which corresponds to the damping rate $\gamma(p)$, since this quantity is of next-to-leading order (NLO) in the HTL formalism. The numerically extracted damping rate is shown in the right panel of \fig\ref{fig_spectral} for both transverse and longitudinal polarizations as a function of momentum for two lattice discretizations. The damping rate is observed to grow for low momenta $p \lesssim 0.15\,\Q \approx m$. Interestingly, both polarizations follow the same curves within their uncertainties. At higher momenta, the curves become flat. The value $\gamma_{\HTL}(p=0)$ calculated in the HTL formalism to NLO \cite{Braaten:1990it} is seen to be roughly consistent with the numerical results.
This shows that the system is dominated by quasiparticles with relatively narrow width. Hence, kinetic theory is indeed applicable to this far-from-equilibrium system. In particular, this is the first time that the full momentum dependence of the damping rates was computed. This extends the known HTL results and hence, improves our understanding also of the thermal state at high temperatures where HTL is applicable.

\section{Conclusion}
\label{sec_conclusion}

We have discussed the emergence and relevance of NTFPs for different physical systems. In ultra-relativistic heavy-ion collisions (URHICs), an NTFP helped to establish our current understanding and weak-coupling description of the dynamics of pre-equilibrated matter created in the collision \cite{Berges:2013eia,Berges:2013fga}. Some important open questions remain to complete our understanding of initial stages in URHICs. One difficulty is the correct description of the instability phase leading from the initial state right after the collision to the onset of kinetic theory. While it is technically challenging to correctly include vacuum fluctuations \cite{Berges:2013lsa,Gelis:2013rba,Berges:2014yta}, the duration and details of this phase are sensitive to the way how the quantum vacuum is modeled as well as to other model parameters in the initial state \cite{Romatschke:2006nk,Philipsen:2018nwr}. Having a detailed description of the earliest stages has important phenomenological consequences, allowing to assess transport coefficients at early times. 

The same scaling behavior emerges in the low-momentum region of scalar theories describing ultra-cold atoms and cosmological models of inflation or of dark matter, forming a broad universality far from equilibrium \cite{Orioli:2015dxa}. Its universal quantities can be reproduced by the Large-$N$ kinetic theory \cite{Orioli:2015dxa,Walz:2017ffj}. Despite the progress in recent years, some aspects of the IR region are not fully understood yet, as, e.g., when quasiparticles or defects dominate for different $N$ \cite{Nowak:2010tm,Orioli:2015dxa,Mikheev:2018adp,Gasenzer:2011by,Moore:2015adu}. 

A promising method to address these problems has been recently developed in \cite{Boguslavski:2018beu} for non-Abelian gauge theories and in \cite{PineiroOrioli:2018hst} for non-relativistic scalar theory. Combining linear response theory with classical simulations, spectral functions can be computed numerically in highly occupied systems. In \cite{Boguslavski:2018beu}, it was applied to a non-Abelian plasma close to its NTFP and it was shown that the results were consistent with HTL theory but also indicated effects surpassing its leading order. Specifically, the momentum dependence of the damping rates was obtained for the first time, which had not been calculated with another method before. By treating the Glasma field and its instabilities numerically separate \cite{Kurkela:2016mhu}, it can be used to study initial stages in URHICs. Extending the method to relativistic scalar theories, also the $N$ dependence of the IR region can be studied.\footnote{K.\ Boguslavski, A.\ Pi\~{n}eiro Orioli, \it{work in progress}.}

\bibliographystyle{JHEP-2modlong}
\bibliography{spires_no_eprint}

\end{document}